\documentclass[12pt]{article}%
\usepackage{amsmath}
\usepackage{amsfonts}
\usepackage{amssymb}
\usepackage{graphicx}%
\providecommand{\U}[1]{\protect\rule{.1in}{.1in}}
\begin{document}
\begin{titlepage}
\ \\
\begin{center}
\LARGE
{\bf
Quantum Energy Teleportation\\
with Electromagnetic Field:\\
Discrete vs. Continuous Variables
}
\end{center}
\ \\
\begin{center}
\large{
Masahiro Hotta
}\\
\ \\
\ \\
{\it
Department of Physics, Faculty of Science, Tohoku University,\\
Sendai 980-8578, Japan\\
hotta@tuhep.phys.tohoku.ac.jp
}
\end{center}
\begin{abstract}
It is well known that usual quantum teleportation protocols cannot transport energy. Recently,  new protocols called quantum
energy teleportation (QET) have been proposed, which transport energy
by local operations and classical  communication
with the ground states of many-body quantum systems.
In this paper, we compare two different QET protocols for transporting
energy with electromagnetic field. In the first protocol,
a 1/2 spin (a qubit) is coupled with the quantum fluctuation in the vacuum state and
measured in order to obtain one-bit information about
the fluctuation for the teleportation. In the second protocol, a harmonic
oscillator is coupled with the fluctuation and measured
in order to obtain continuous-variable information about
the fluctuation. In the spin protocol,
the amount of teleported energy is suppressed
by an exponential damping factor
when the amount of input energy increases.
This suppression factor becomes power damping
in the case of the harmonic oscillator protocol.
Therefore, it is concluded that obtaining more information about the quantum fluctuation
leads to  teleporting more energy. This result suggests a profound relationship between energy and quantum information.
\end{abstract}
\end{titlepage}

\bigskip

\section{Introduction}

\ \newline

In quantum field theory, the concept of negative energy physics has attracted
considerable attention for a long time. Quantum interference can produce
various states containing regions of negative energy, although the total
energy remains nonnegative \cite{BD}. The concept of negative energy plays
important roles in many fundamental problems of physics, including traversable
wormhole \cite{wt}, cosmic censorship \cite{cc}, and the second law of
thermodynamics \cite{Ford}. In addition, its physical application to quantum
optics has been discussed \cite{optics}. Recently, negative energy physics has
yielded a quantum protocol called quantum energy teleportation (QET) in which
energy can be transported using only local operations and classical
communication (LOCC) without breaking causality and local energy conservation
\cite{hotta1}-\cite{hotta-ion}. QET can be theoretically considered in various
many-body quantum systems including 1+1 dimensional massless fields
\cite{hotta1}, spin chains \cite{hotta2} and cold trapped ions
\cite{hotta-ion}.  Based on developing measurement technology with sensitive
energy resolution for the systems, the QET effect might be observable in
future. It may be also possible to enhance the amount of teleported energy by
preparing a large number of parallel QET channels, performing a QET protocol
for each channel and accumulating each teleported energy so as to achieve
desired amount of total energy. After future experimental verification of QET,
amazing possibility would be open in principle for nano-technology application
of QET. For example, it may be imagined that, without heat generation in the
intermediate subsystems of the QET channels, energy is transported in
nano-machines at a speed much faster than the evolution speed of excitations
of the channels. This technology, if possible, helps future development of
quantum computers in which energy distribution and quantum tasks in the
devices are completed before heat generation in the system. QET is also
expected to provide insights on unsolved problems in gravitational physics. In
fact, a QET process has already been analyzed in black hole physics, and from
the measured information of zero-point oscillation of quantum fields, it can
be regarded as controlled black hole evaporation if we consider the protocol
near the horizon of a large-mass black hole \cite{hotta3}.

Energy transportation usually requires physical carriers of energy such as
electric currents and radiation waves. Energy is infused into the gateway
point of a transport channel connected to a distant exit point. Then, energy
carriers of the channel excite and propagate to the exit point. At the exit
point, energy is extracted from the carriers and harnessed for many purposes.
On the other hand, in the QET protocols, energy can be extracted from the exit
point even if no excited energy carriers arrive at the exit point of the
channel. We locally measure quantum fluctuation around the gateway point in
the ground state of the channel system and announce the measurement result to
the distant exit point with zero energy density, where we can extract energy
from the channel. A key feature is that this measurement result includes
information about the quantum fluctuation of the channel around this distant
point via quantum correlation of the ground state of the channel system.
Therefore, we can infer details about the behavior of a distant fluctuation
from the result of the local measurement. To compensate the extraction of this
information, some amount of energy must be infused into the  channel system at
the measurement point; this is regarded as input energy to the gateway point
of the channel. By choosing and performing a proper local operation based on
the announced information at the distant point, the local zero-point
oscillation around the distant point can be suppressed relative to the
ground-state one, yielding a negative energy density. During the operation,
respecting local energy conservation, positive amount of surplus energy is
moved from the channel system to external systems. This is regarded as output
teleported energy from the exit point of the channel.

One of the important unresolved problems in QET is the theoretical
clarification of the properties in 1+3 dimensions. Protocols in 1+1 dimensions
have already been extensively analyzed in\ previous studies \cite{hotta1}%
-\cite{hotta-ion}. However, 1+3 dimensional models have not yet been analyzed.
In addition, all the protocols proposed thus far adopt quantum measurements
for discrete-variable information. Therefore, it would be interesting to
investigate not only a protocol with discrete-variable information but also
one with continuous-variable information. In this study, we carry out a
detailed analysis of two QET protocols for 1+3 dimensional electromagnetic
field in the Coulomb gauge. Local measurements of quantum fluctuations in the
vacuum state of the field require energy infusion to the field. The infused
energy is diffused to spatial infinity at the velocity of light and the state
of the field soon becomes a local vacuum with zero energy around the
measurement area. Obviously, this escaped energy cannot be taken back to the
measurement area by local operations around this area if we do not know the
measurement result of the fluctuation. However, if the measurement result is
available, we can effectively take back a part of this energy to the
measurement area by applying the QET mechanism. By carrying out a local
unitary operation dependent on the measurement result for the measurement area
with zero energy density, the fluctuation of zero-point oscillation is
squeezed and a negative energy density appears around the area, accompanied by
the extraction of positive energy from the fluctuation to external systems.
Needless to say, without the measurement result, it is impossible to extract
energy from the zero-energy fluctuation. One of the two QET protocols we will
consider is a teleportation in which discrete-variable information about a
fluctuation is obtained using a measurement with a 1/2 spin (a qubit), and the
other is a teleportation in which continuous-variable information is obtained
using a measurement with a harmonic oscillator. The discrete-variable protocol
is a straightforward extension of the protocol for a 1+1 dimensional field
proposed in \cite{hotta1}. The measurements are generalized (POVM) ones that
use probe systems (1/2 spin and harmonic oscillator) strongly interacting with
local electric field fluctuations during a short time. We prove that for a
large energy input, the continuous-variable teleportation is more attractive
than the discrete-variable teleportation. In the discrete-variable case, the
amount of teleported energy is suppressed by an exponential damping factor
when the amount of energy infused by the measurement increases. Meanwhile,
this suppression factor becomes power damping in the continuous-variable case.
Therefore, it is concluded that obtaining more information about the quantum
fluctuation leads to teleporting more energy. This result suggests a new
profound relation between energy and quantum information. So far, relationship
between energy and information has been extensively discussed only in the
context of computation energy cost \cite{L}, \cite{B}, \cite{su}. The QET
viewpoint may shed light on a new  relationship between amount of teleported
energy and amount of quantum information about ground-state fluctuations which
would be characterized by various informational indices including mutual
information and entanglement. The explicit analysis about this  relationship
is beyond the scope of this paper and will be reported elsewhere.

The remainder of this paper is organized as follows. In section 2, a brief
review of the quantization of the electromagnetic field in the Coulomb gauge
is presented in order to clarify our notations, and the emergence of negative
energy density is explained. In section 3, we discuss a discrete-variable
protocol. In section 4, a continuous-variable protocol is analyzed. In section
5, a summary and discussions are presented. In this paper, we adopt the
natural unit\bigskip\ $c=\hbar=1$.

\section{Quantization in Coulomb Gauge}

\ \newline

We present a short review of quantization of the electromagnetic field in the
Coulomb gauge in order to clarify the notations used for later discussions.
The gauge is defined by%
\[
A_{0}=0,~\operatorname{div}\mathbf{A}=0
\]
for the gauge field $A_{\mu}$ =$\left(  A_{0},\mathbf{A}\right)  $. Then, the
equation of motion of the Heisenberg operator of the gauge field
$\mathbf{\hat{A}}(t,\mathbf{x})$ is reduced to the massless Klein-Gordon
equation given by%
\[
\left[  \partial_{t}^{2}-\nabla^{2}\right]  \mathbf{\hat{A}}=0.
\]
The solution can be expanded in terms of plain-wave modes as follows.%
\[
\mathbf{\hat{A}}(t,\mathbf{x})=\int\frac{d^{3}k}{\sqrt{\left(  2\pi\right)
^{3}2|\mathbf{k}|}}\sum_{h=1,2}\left[  \mathbf{e}_{h}\left(  \mathbf{k}%
\right)  \hat{a}_{\mathbf{k}}^{h}e^{i\left(  \mathbf{k}\cdot\mathbf{x}%
-|\mathbf{k}|t\right)  }+\mathbf{e}_{h}^{\ast}\left(  \mathbf{k}\right)
\hat{a}_{\mathbf{k}}^{h\dag}e^{-i\left(  \mathbf{k}\cdot\mathbf{x}%
-|\mathbf{k}|t\right)  }\right]  ,
\]
where $~\hat{a}_{\mathbf{k}}^{h\dag}$ $\left(  \hat{a}_{\mathbf{k}}%
^{h}\right)  $ is a creation (annihilation) operator of the photon with
momentum $\mathbf{k}$ and polarization $h$ satisfying
\[
\left[  \hat{a}_{\mathbf{k}}^{h},~\hat{a}_{\mathbf{k}^{\prime}}^{h^{\prime
}\dag}\right]  =\delta_{hh^{\prime}}\delta\left(  \mathbf{k}-\mathbf{k}%
^{\prime}\right)  ,
\]
and $\mathbf{e}_{h}\left(  \mathbf{k}\right)  $ is a polarization vector
satisfying
\begin{align*}
\mathbf{e}_{h}\left(  \mathbf{k}\right)  ^{\ast}\cdot\mathbf{e}_{h^{\prime}%
}\left(  \mathbf{k}\right)   &  =\delta_{hh^{\prime}},\\
\mathbf{k}\cdot\mathbf{e}_{h}\left(  \mathbf{k}\right)   &  =0.
\end{align*}
In this study, because we take a sum of two polarization contributions to
obtain the final results, the reality condition can be imposed on
$\mathbf{e}_{h}\left(  \mathbf{k}\right)  $ for simplicity such that
$\mathbf{e}_{h}^{\ast}\left(  \mathbf{k}\right)  =\mathbf{e}_{h}\left(
\mathbf{k}\right)  $. In addition, $\mathbf{e}_{h}\left(  \mathbf{k}\right)  $
satisfies the completeness relation as%

\[
\sum_{h}e_{h}^{a}\left(  \mathbf{k}\right)  e_{h}^{b}\left(  \mathbf{k}%
\right)  =\delta_{ab}-\frac{k_{a}k_{b}}{\mathbf{k}^{2}}.
\]
The energy density operator of the field is defined by%

\[
\hat{\varepsilon}\left(  \mathbf{x}\right)  =\frac{1}{2}:\left(
\mathbf{\hat{E}}\left(  \mathbf{x}\right)  ^{2}+\left(  \nabla\times
\mathbf{\hat{A}}\left(  \mathbf{x}\right)  \right)  ^{2}\right)  :,
\]
where $\mathbf{\hat{E}}\left(  \mathbf{x}\right)  $ is the electric field
operator and $::$ denotes the standard normal order with respect to $\hat
{a}_{\mathbf{k}}^{h}$ and$~\hat{a}_{\mathbf{k}}^{h\dag}$. The Hamiltonian is
given by the spatial integration of the energy density as follows.%

\[
\hat{H}=\int\hat{\varepsilon}\left(  \mathbf{x}\right)  d^{3}x.
\]
It is a well-known fact that the Hamiltonian (total energy of the field) is a
nonnegative operator. The vacuum state $|0\rangle$ is the eigenstate with the
lowest eigenvalue zero of $\hat{H}$ as $\hat{H}|0\rangle=0.$ The expectation
value of energy density vanishes for the vacuum state as%
\begin{equation}
\langle0|\hat{\varepsilon}\left(  \mathbf{x}\right)  |0\rangle=0. \label{72}%
\end{equation}

In the later discussion, coherent states are used often. Therefore, we present
a summary of the related properties of the coherent states. A displacement
operator generating a coherent state from $|0\rangle$ is given by%
\[
\hat{U}\left(  \mathbf{p,q}\right)  =\exp\left[  i\int\left[  \mathbf{p}%
(\mathbf{x})\cdot\mathbf{\hat{A}}(\mathbf{x})-\mathbf{q}(\mathbf{x}%
)\cdot\mathbf{\hat{E}(\mathbf{x})}\right]  d^{3}x\right]  ,
\]
where $\mathbf{p}(\mathbf{x})$ and $\mathbf{q}(\mathbf{x})$ are real vector
functions satisfying the conditions of this gauge as
\[
\nabla\cdot\mathbf{p}(\mathbf{x})=0,
\]%
\[
\nabla\cdot\mathbf{q}(\mathbf{x})=0.
\]
By using the commutation relation between the gauge field and the electric
field at time $t=0$ given by%

\[
\left[  \hat{A}_{a}(\mathbf{x}),~\hat{E}_{b}(\mathbf{y})\right]  =i\left(
\delta_{ab}-\frac{\partial_{a}\partial_{b}}{\nabla^{2}}\right)  \delta\left(
\mathbf{x}-\mathbf{y}\right)  ,
\]
it is easily verified that $\hat{U}\left(  \mathbf{p,q}\right)  $ displaces
$\mathbf{\hat{A}}(\mathbf{x})$ and $\mathbf{\hat{E}}(\mathbf{x})$ as%

\begin{equation}
\hat{U}^{\dag}\left(  \mathbf{p,q}\right)  \mathbf{\hat{E}}(\mathbf{x})\hat
{U}\left(  \mathbf{p,q}\right)  =\mathbf{\hat{E}}(\mathbf{x})+\mathbf{p}%
(\mathbf{x}), \label{e5}%
\end{equation}

\begin{equation}
\hat{U}^{\dag}\left(  \mathbf{p,q}\right)  \mathbf{\hat{A}}(\mathbf{x})\hat
{U}\left(  \mathbf{p,q}\right)  =\mathbf{\hat{A}}(\mathbf{x})+\mathbf{q}%
(\mathbf{x}). \label{e6}%
\end{equation}
In addition, we are able to prove a product formula as
\begin{align*}
&  \hat{U}\left(  \mathbf{p}_{1}\mathbf{,q}_{1}\right)  \hat{U}\left(
\mathbf{p}_{2}\mathbf{,q}_{2}\right) \\
&  =\exp\left[  \frac{i}{2}\int\left(  \mathbf{p}_{1}\mathbf{\cdot q}%
_{2}\mathbf{-q}_{1}\mathbf{\cdot p}_{2}\right)  d^{3}x\right]  \hat{U}\left(
\mathbf{p}_{1}+\mathbf{p}_{2}\mathbf{,q}_{1}+\mathbf{q}_{2}\right)  .
\end{align*}
This implies that the set of $\hat{U}\left(  \mathbf{p,q}\right)  $ forms a
unitary ray representation of the displacement group of the field. Coherent
states generated by $\hat{U}\left(  \mathbf{p,q}\right)  $ are defined by
\begin{equation}
|\left(  \mathbf{p,q}\right)  \rangle=\hat{U}\left(  \mathbf{p,q}\right)
|0\rangle. \label{e4}%
\end{equation}
By using the Fourier transformation of $\mathbf{p}(\mathbf{x})$ and
$\mathbf{q}(\mathbf{x})$ defined by%

\begin{align*}
\mathbf{P}\left(  \mathbf{k}\right)   &  =\int\mathbf{p}(\mathbf{x}%
)e^{-i\mathbf{k}\cdot\mathbf{x}}d^{3}x,\\
\mathbf{Q}\left(  \mathbf{k}\right)   &  =\int\mathbf{q}(\mathbf{x}%
)e^{-i\mathbf{k}\cdot\mathbf{x}}d^{3}x,
\end{align*}
the coherent states are explicitly written in terms of the creation operator
$\hat{a}_{\mathbf{k}}^{h\dag}$ as follows.%

\begin{align*}
&  |\left(  \mathbf{p,q}\right)  \rangle\\
&  =\exp\left[  -\frac{1}{2}\int\frac{d^{3}k}{\left(  2\pi\right)
^{3}2|\mathbf{k}|}\left\vert \mathbf{P}\left(  \mathbf{k}\right)
-i|\mathbf{k}|\mathbf{Q}\left(  \mathbf{k}\right)  \right\vert ^{2}\right] \\
&  \times\exp\left[  i\int\frac{d^{3}k}{\sqrt{\left(  2\pi\right)
^{3}2|\mathbf{k}|}}\sum_{h}\mathbf{e}_{h}\left(  \mathbf{k}\right)
\cdot\left(  \mathbf{P}\left(  \mathbf{k}\right)  -i|\mathbf{k}|\mathbf{Q}%
\left(  \mathbf{k}\right)  \right)  a_{\mathbf{k}}^{h\dag}\right]  |0\rangle.
\end{align*}
From the above expression, it is easy to prove that $|\left(  \mathbf{p,q}%
\right)  \rangle$ is an eigenstate of the annihilation operator $\hat
{a}_{\mathbf{k}}^{h}$ such that%

\begin{equation}
\hat{a}_{\mathbf{k}}^{h}|\left(  \mathbf{p,q}\right)  \rangle=\frac{i}%
{\sqrt{\left(  2\pi\right)  ^{3}2|\mathbf{k}|}}\mathbf{e}_{h}\left(
\mathbf{k}\right)  \cdot\left(  \mathbf{P}\left(  \mathbf{k}\right)
-i|\mathbf{k}|\mathbf{Q}\left(  \mathbf{k}\right)  \right)  |\left(
\mathbf{p,q}\right)  \rangle. \label{e22}%
\end{equation}
The inner product of two coherent states is explicitly calculated as%
\begin{align}
&  \langle\left(  \mathbf{p}_{1}\mathbf{,q}_{1}\right)  \mathbf{|}\left(
\mathbf{p}_{2}\mathbf{,q}_{2}\right)  \rangle\nonumber\\
&  =e^{\frac{i}{2}\int\left(  \mathbf{p}_{1}\mathbf{\cdot q}_{2}%
\mathbf{-q}_{1}\mathbf{\cdot p}_{2}\right)  d^{3}x}\nonumber\\
&  \times e^{-\frac{1}{2}\int\frac{d^{3}k}{\left(  2\pi\right)  ^{3}%
2|\mathbf{k}|}\left\vert \mathbf{P}_{1}\left(  \mathbf{k}\right)
-\mathbf{P}_{2}\left(  \mathbf{k}\right)  -i|\mathbf{k}|\left(  \mathbf{Q}%
_{1}\left(  \mathbf{k}\right)  -\mathbf{Q}_{2}\left(  \mathbf{k}\right)
\right)  \right\vert ^{2}}. \label{100}%
\end{align}

Next, we examine the emergence of a region with negative energy density in
this standard theory. As a simple example \cite{Ford}, let us consider a
superposition state $|\Psi\rangle$ of the vacuum state $|0\rangle$ and a
two-photon state $|2\rangle$ such that%

\[
|\Psi\rangle=\cos\theta|0\rangle+e^{i\delta}\sin\theta|2\rangle,
\]
where $\theta$ and $\delta$ are real parameters with $\theta\in\left[
0,\pi\right]  $ and $\delta\in\left[  0,2\pi\right]  $. Generally, an
off-diagonal element of the energy density $\langle0|\hat{\varepsilon}\left(
\mathbf{x}\right)  |2\rangle$ does not vanish for a fixed point $\mathbf{x}$.
This is because $\hat{\varepsilon}\left(  \mathbf{x}\right)  $ includes a
non-vanishing term proportional to $\hat{a}_{\mathbf{k}}^{h}\hat
{a}_{\mathbf{k}^{\prime}}^{h^{\prime}}$. This fact indicates the emergence of
negativity of the energy density as follows. The expectation value of the
energy density for the state $|\Psi\rangle$ is calculated as%

\begin{align*}
&  \langle\Psi|\hat{\varepsilon}\left(  \mathbf{x}\right)  |\Psi\rangle\\
&  =2\cos\theta\sin\theta\left(  \cos\delta\operatorname{Re}\langle
0|\hat{\varepsilon}\left(  \mathbf{x}\right)  |2\rangle-\sin\delta
\operatorname{Im}\langle0|\hat{\varepsilon}\left(  \mathbf{x}\right)
|2\rangle\right) \\
&  +\sin^{2}\theta\langle2|\hat{\varepsilon}\left(  \mathbf{x}\right)
|2\rangle.
\end{align*}
In this result, let us set the parameters $\theta$ and $\delta$ so as to
satisfy
\begin{align*}
\cos\theta &  =\frac{\langle2|\hat{\varepsilon}\left(  \mathbf{x}\right)
|2\rangle}{\sqrt{\langle2|\hat{\varepsilon}\left(  \mathbf{x}\right)
|2\rangle^{2}+4\left\vert \langle0|\hat{\varepsilon}\left(  \mathbf{x}\right)
|2\rangle\right\vert ^{2}}},\\
\sin\theta &  =\frac{2\left\vert \langle0|\hat{\varepsilon}\left(
\mathbf{x}\right)  |2\rangle\right\vert }{\sqrt{\langle2|\hat{\varepsilon
}\left(  \mathbf{x}\right)  |2\rangle^{2}+4\left\vert \langle0|\hat
{\varepsilon}\left(  \mathbf{x}\right)  |2\rangle\right\vert ^{2}}},\\
\tan\delta &  =-\frac{\operatorname{Im}\langle0|\hat{\varepsilon}\left(
\mathbf{x}\right)  |2\rangle}{\operatorname{Re}\langle0|\hat{\varepsilon
}\left(  \mathbf{x}\right)  |2\rangle}.
\end{align*}
Then, $\langle\Psi|\hat{\varepsilon}\left(  \mathbf{x}\right)  |\Psi\rangle$
is evaluated as a negative value as follows.%

\[
\langle\Psi|\hat{\varepsilon}\left(  \mathbf{x}\right)  |\Psi\rangle=-\frac
{1}{2}\left[  \sqrt{\langle2|\hat{\varepsilon}\left(  \mathbf{x}\right)
|2\rangle^{2}+4\left\vert \langle0|\hat{\varepsilon}\left(  \mathbf{x}\right)
|2\rangle\right\vert ^{2}}-\langle2|\hat{\varepsilon}\left(  \mathbf{x}%
\right)  |2\rangle\right]  <0.
\]
Therefore, the emergence of negative-energy regions is not a peculiar
phenomenon in quantum field theory. Quantum interference in the superposition
of photon number eigenstates yields negative values. It should be
re-emphasized that despite the existence of regions with negative energy
density, the expectation values of $\hat{H}$ remain nonnegative. This implies
that there exist regions with a sufficient amount of positive energy so as to
make the total energy nonnegative. As described in sections 3 and 4, this
negative energy plays a crucial role in the QET protocols.

\bigskip

\section{Discrete-Variable Teleportation}

\bigskip

Our protocol for QET with a 1/2 spin probe is a straightforward extension of
that in \cite{hotta1} and comprises the following three steps.

\bigskip

(1) At time $t=0$, the spin probe is strongly coupled with the vacuum
fluctuation of the electric field within a finite compact region $V_{m}$
during a very short time. In this process, some information about the
fluctuation is imprinted into the probe. Positive energy is inevitably infused
into the field during the measurement process, as seen later. The amount of
energy is denoted by $E_{m}$. When this energy is infused, positive-energy
wave packets of the field are generated and these propagate to spatial
infinity with the velocity of light.

\bigskip

(2) After switching off the interaction, projective measurement of the
$z$-component of the spin is carried out. If the up or down state is observed,
we assign $s=+$ or $-$, respectively, to the measurement result. This implies
that we obtains one-bit information about the field fluctuation via the probe measurement.

\bigskip

(3) At time $t=T$, it is assumed that the measurement has finished and the
wave packets have already escaped from the region. Hence, the energy density
in the region $V_{m}$ is exactly zero. Then, a local displacement operation is
carried out depending on $s$ within $V_{m}$. Even though we have zero energy
in $V_{m}$, positive energy is extracted from the field fluctuation during the
local operation, generating negative-energy wave packets of the field. The
amount of negative energy of the wave packets is denoted by $E_{o}%
(=-\left\vert E_{o}\right\vert )$. Therefore, the amount of energy extracted
energy from the field is given by $+\left\vert E_{o}\right\vert $.

\bigskip

In step (1), the measurement interaction between the electric field and the
spin probe is given by%

\begin{equation}
\hat{H}_{m}(t)=g(t)\hat{\sigma}_{z}\hat{G}, \label{e20}%
\end{equation}
where $\hat{\sigma}_{z}$ is the $z$-component of the Pauli matrices of the
spin probe; $g(t)$, a time-dependent real coupling constant; and $\hat{G}$, a
Hermitian operator defined by%

\begin{equation}
\hat{G}=\frac{\pi}{4}-\int\mathbf{a}_{m}(\mathbf{x})\cdot\mathbf{\hat{E}%
}(\mathbf{x})d^{3}x. \label{81}%
\end{equation}
Here, $\mathbf{a}_{m}(\mathbf{x})$ is a real vector function with a support
equal to $V_{m}$ satisfying $\nabla\cdot\mathbf{a}_{m}(\mathbf{x})=0$. In
addition, by taking a short-time limit for switching the interaction, we set%

\begin{equation}
g(t)=\delta(t-0), \label{e1}%
\end{equation}
The initial state of the spin probe is assumed to be the up state
$|+_{x}\rangle$ of the $x$-component of the spin given by%
\[
|+_{x}\rangle=\frac{1}{\sqrt{2}}\left[
\begin{array}
[c]{c}%
1\\
1
\end{array}
\right]  .
\]
In step (2), the measurement operators $\hat{M}_{\pm}$ \cite{nc} are defined by%

\begin{equation}
\hat{M}_{\pm}=\langle\pm_{x}|\operatorname*{T}\exp\left[  -i\int_{0}^{t_{m}%
}\hat{H}_{m}(t)dt\right]  |+_{x}\rangle, \label{60}%
\end{equation}
where $|-_{x}\rangle$ is the down state of the $x$-component of the spin given
by
\[
|-_{x}\rangle=\frac{-i}{\sqrt{2}}\left[
\begin{array}
[c]{c}%
1\\
-1
\end{array}
\right]  .
\]
By using Eq. (\ref{e1}), $\hat{M}_{\pm}$ are computed as%

\begin{align}
\hat{M}_{+}  &  =\cos\hat{G},\label{e62}\\
\hat{M}_{-}  &  =\sin\hat{G}. \label{e63}%
\end{align}
Using Eq. (\ref{e4}), the post-measurement states of the field obtaining
$s=\pm$ are calculated as%

\begin{align*}
|\psi_{+}\rangle &  =\frac{1}{\sqrt{p_{+}}}\hat{M}_{+}|0\rangle=\frac
{1}{2\sqrt{p_{+}}}\left[  e^{i\frac{\pi}{4}}|\left(  \mathbf{0,a}_{m}\right)
\rangle+e^{-i\frac{\pi}{4}}|\left(  \mathbf{0,-a}_{m}\right)  \rangle\right]
,\\
|\psi_{-}\rangle &  =\frac{1}{\sqrt{p_{-}}}\hat{M}_{-}|0\rangle=\frac
{1}{2i\sqrt{p_{-}}}\left[  e^{i\frac{\pi}{4}}|\left(  \mathbf{0,a}_{m}\right)
\mathbf{\rangle}-e^{-i\frac{\pi}{4}}|\left(  \mathbf{0,}-\mathbf{a}%
_{m}\right)  \rangle\right]  ,
\end{align*}
where $p_{\pm}$ is the measurement probability of $|\pm_{x}\rangle$ for the
spin probe and it is evaluated as
\[
p_{\pm}=\langle0|\hat{M}_{\pm}^{\dag}\hat{M}_{\pm}|0\rangle=\frac{1}{2}.
\]
After the measurement, the average state of the field is given by
\[
\hat{\rho}_{\hat{M}}=\sum_{s=\pm}p_{s}|\psi_{s}\rangle\langle\psi_{s}|.
\]
The amount of energy $E_{m}$ of the field after the measurement is not zero
but is instead a positive value given by
\begin{equation}
E_{m}=\int\operatorname*{Tr}\left[  \hat{\rho}_{\hat{M}}\hat{\varepsilon
}(\mathbf{x})\right]  d^{3}x=\frac{1}{2}\int\left(  \nabla\times\mathbf{a}%
_{m}(\mathbf{x)}\right)  ^{2}d^{3}x. \label{e7}%
\end{equation}
This evaluation of Eq. (\ref{e7}) is achieved by using Eq. (\ref{e22}).
Because the initial state of the field is the vacuum state and it has zero
energy, the positive value in Eq. (\ref{e7}) implies that the manipulation of
the measurement requires positive work from outside. In general, any local
operation on the vacuum state infuses nonzero energy into the field if the
post-operation state is not the vacuum state, because the Hamiltonian is a
nonnegative operator. In Figure 1, the measurement process of QET (step (1))
is depicted in the $xy$-plane slice. The measurement area $V_{m}$ is
represented by a circle in the plane.

Next, let us discuss time evolution of the field after the measurement. The
average state evolves as%

\[
\hat{\rho}_{m}(t)=\sum_{s=\pm}p_{s}\hat{U}(t)|\psi_{s}\rangle\langle\psi
_{s}|\hat{U}(t)^{\dag}%
\]
by the time-evolution operator $\hat{U}(t)=\exp\left[  -it\hat{H}\right]  $.
The Heisenberg operators $\mathbf{\hat{A}}(t,\mathbf{x})$ and $\mathbf{\hat
{E}}(t,\mathbf{x})$ are given by the Schr\H{o}dinger operators $\mathbf{\hat
{A}}(\mathbf{x})\left(  =\mathbf{\hat{A}}(0,\mathbf{x})\right)  $ and
$\mathbf{\hat{E}}(\mathbf{x})\left(  =\mathbf{\hat{E}}(0,\mathbf{x})\right)  $
as follows.%
\begin{align}
\mathbf{\hat{A}}(t,\mathbf{x})  &  =\int\Delta^{(2)}(t,\mathbf{x}%
-\mathbf{y})\mathbf{\hat{A}}(0,\mathbf{y})d^{3}y\nonumber\\
&  +\int\Delta^{(1)}(t,\mathbf{x}-\mathbf{y})\mathbf{\hat{E}}(0,\mathbf{y}%
)d^{3}y, \label{e8}%
\end{align}%
\begin{align}
\mathbf{\hat{E}}(t,\mathbf{x})  &  =\int\partial_{t}\Delta^{(2)}%
(t,\mathbf{x}-\mathbf{y})\mathbf{\hat{A}}(0,\mathbf{y})d^{3}y\nonumber\\
&  +\int\Delta^{(2)}(t,\mathbf{x}-\mathbf{y})\mathbf{\hat{E}}(0,\mathbf{y}%
)d^{3}y, \label{e9}%
\end{align}
where $\Delta^{(1)}$ and $\Delta^{(2)}$ are Lorentz-invariant functions
defined by%

\begin{align}
\Delta^{(1)}(t,\mathbf{x}-\mathbf{y})  &  =\frac{1}{2\pi}\epsilon
(t)\delta\left(  t^{2}-\left\vert \mathbf{x}-\mathbf{y}\right\vert
^{2}\right)  ,\nonumber\\
\Delta^{(2)}(t,\mathbf{x}-\mathbf{y})  &  =\partial_{t}\Delta^{(1)}%
(t,\mathbf{x}-\mathbf{y}). \label{900}%
\end{align}
Both the functions $\Delta^{(1)}$ and $\Delta^{(2)}$ vanish in the
off-light-cone region with $t^{2}-\left\vert \mathbf{x}-\mathbf{y}\right\vert
^{2}$ $\neq0$ and satisfy the massless Klein-Gordon equation:%

\[
\left[  \partial_{t}^{2}-\nabla^{2}\right]  \Delta^{(n)}(t,\mathbf{x}%
-\mathbf{y})=0.
\]
Substituting Eq. (\ref{e8}) and Eq. (\ref{e9}) into $\hat{\varepsilon
}(t,\mathbf{x})$ yields the time evolution of the average value of the energy
density as%

\begin{equation}
\langle\hat{\varepsilon}(t,\mathbf{x})\rangle=\operatorname*{Tr}\left[
\hat{\rho}_{m}(t)\hat{\varepsilon}(\mathbf{x})\right]  =\frac{1}{2}\left[
\mathbf{\Pi}(t,\mathbf{x)}^{2}+\mathbf{b}(t,\mathbf{x)}^{2}\right]  ,
\label{e16}%
\end{equation}
where $\mathbf{\Pi}(t,\mathbf{x)}$ and $\mathbf{b}(t,\mathbf{x)}$ are given by%

\begin{align}
\mathbf{b}(t,\mathbf{x)}  &  \mathbf{=}\int\Delta^{(2)}(t,\mathbf{x}%
-\mathbf{y})\nabla\times\mathbf{a}_{m}(0,\mathbf{y})d^{3}y,\label{e10}\\
\mathbf{\Pi}(t,\mathbf{x)}  &  \mathbf{=}\int\partial_{t}\Delta^{(2)}%
(t,\mathbf{x}-\mathbf{y})\mathbf{a}_{m}(0,\mathbf{y})d^{3}y. \label{e12}%
\end{align}
Taking account of the explicit form of $\Delta^{(2)}(t,\mathbf{x}-\mathbf{y})$
in Eq. (\ref{900}), the above result teaches us that the wave packets excited
by the measurement soon escape from the measurement area to spatial infinity
with the velocity of light. This ensures that the energy density in $V_{m}$
returns to zero. At time $t=T$, the state $\hat{\rho}_{m}(T)$ is assumed to be
a local vacuum state with zero energy density:%

\[
\operatorname*{Tr}\left[  \hat{\rho}_{m}(T)\hat{\varepsilon}(\mathbf{x}%
)\right]  =0,
\]
for $\mathbf{x}\mathbf{\in}V_{m}$.

In step (3), let us consider a local displacement operation within $V_{m}$
depending on $s$ defined by
\begin{equation}
\hat{U}_{s}=\exp\left[  is\theta\int\mathbf{f}_{o}(\mathbf{x})\cdot
\mathbf{\hat{A}}(\mathbf{x})d^{3}x\right]  \label{57}%
\end{equation}
where $\theta$ is a real constant fixed below, and $\mathbf{f}_{o}%
(\mathbf{x})$ is a real vector function with the support $V_{m}$ that
satisfies
\[
\nabla\cdot\mathbf{f}_{o}(\mathbf{x})=0
\]
so as to preserve gauge invariance of $\hat{U}_{s}$. In this process, positive
energy is released on average from the field to the apparatus executing
$\hat{U}_{s}$ by taking a proper value of $\theta$. Let us introduce the
localized energy operator around $V_{m}$ as
\[
\hat{H}_{o}=\int\mathbf{w}(\mathbf{x})\hat{\varepsilon}\left(  \mathbf{x}%
\right)  d^{3}x,
\]
where $\mathbf{w}(\mathbf{x})$ is a real window function of $V_{m}$ that
satisfies $\mathbf{w}(\mathbf{x})=1$ for $\mathbf{x}\mathbf{\in}V_{m}$ and
decays rapidly outside $V_{m}$. The average state after the displacement
operation is given by%

\[
\hat{\rho}=\sum_{s=\pm}\hat{U}_{s}\hat{U}(T)\hat{M}_{s}|0\rangle\langle
0|\hat{M}_{s}^{\dag}\hat{U}^{\dag}(T)\hat{U}_{s}^{\dag}.
\]
For this state, the average energy of the field around $V_{m}$ is defined by%
\begin{equation}
E_{o}=\operatorname*{Tr}\left[  \hat{H}_{o}\hat{\rho}\right]  =\sum_{s=\pm
}\langle0|\hat{M}_{s}^{\dag}\hat{U}^{\dag}(T)\hat{U}_{s}^{\dag}\hat{H}_{o}%
\hat{U}_{s}\hat{U}(T)\hat{M}_{s}|0\rangle. \label{30}%
\end{equation}
Here, the operator $\hat{U}^{\dag}(T)\hat{U}_{s}^{\dag}\hat{H}_{o}\hat{U}%
_{s}\hat{U}(T)$ is rewritten using Eq. (\ref{e5}) and Eq. (\ref{e6}) as%

\begin{align*}
&  \hat{U}^{\dag}(T)\hat{U}_{s}^{\dag}\hat{H}_{o}\hat{U}_{s}\hat{U}(T)\\
&  =\int\mathbf{w}(\mathbf{x})\hat{\varepsilon}\left(  T,\mathbf{x}\right)
d^{3}x\\
&  +s\theta\int\mathbf{f}_{o}\mathbf{(\mathbf{x})}\cdot\mathbf{\hat
{E}\mathbf{(}}T\mathbf{\mathbf{,\mathbf{x})}}d^{3}x+\frac{1}{2}\theta^{2}%
\int\mathbf{f}_{o}\mathbf{(\mathbf{x})}^{2}d^{3}x,
\end{align*}
where $\hat{\varepsilon}\left(  T,\mathbf{x}\right)  $ is the Heisenberg
operator given by $\hat{U}^{\dag}(T)\hat{\varepsilon}\left(  \mathbf{x}%
\right)  \hat{U}(T),$ and we have used $\mathbf{w}(\mathbf{x})\mathbf{f}%
_{o}\mathbf{(\mathbf{x})=f}_{o}\mathbf{(\mathbf{x})}$. Substituting the above
relation into Eq. (\ref{30}) yields the following relation.
\begin{align}
&  E_{o}=\sum_{s=\pm}\langle0|\hat{M}_{s}^{\dag}\left[  \int\mathbf{w}%
(\mathbf{x})\hat{\varepsilon}\left(  T,\mathbf{x}\right)  d^{3}x\right]
\hat{M}_{s}|0\rangle\nonumber\\
&  +\theta\sum_{s=\pm}s\langle0|\hat{M}_{s}^{\dag}\left[  \int\mathbf{f}%
_{o}\mathbf{(\mathbf{x})}\cdot\mathbf{\hat{E}\mathbf{(}}%
T\mathbf{\mathbf{,\mathbf{x})}}d^{3}x\right]  \hat{M}_{s}|0\rangle\nonumber\\
&  +\frac{1}{2}\theta^{2}\int\mathbf{f}_{o}\mathbf{(\mathbf{x})}^{2}%
d^{3}x\langle0|\left(  \sum_{s=\pm}\hat{M}_{s}^{\dag}\hat{M}_{s}\right)
|0\rangle. \label{31}%
\end{align}
In order to further simplify the form of $E_{o}$, the following equation is used.%

\begin{align}
&  \left[  \int\mathbf{f}_{o}\mathbf{(\mathbf{x})}\cdot\mathbf{\hat
{E}\mathbf{(}}T\mathbf{\mathbf{,\mathbf{x})}}d^{3}x,~\int\mathbf{a}%
_{m}(\mathbf{y})\cdot\mathbf{\hat{E}}(0,\mathbf{y})d^{3}y\right] \nonumber\\
&  =i\int\int\mathbf{f}_{o}\mathbf{(\mathbf{x})}\cdot\mathbf{a}_{m}%
(\mathbf{y})\partial_{T}\Delta^{(2)}(T,\mathbf{x}-\mathbf{y})d^{3}xd^{3}y
\label{32}%
\end{align}
This is proven by Eq. (\ref{e9}). Taking account of the supports of
$\mathbf{f}_{o}\mathbf{(\mathbf{x})}$ and $\mathbf{a}_{m}(\mathbf{y})$, it is
verified that the integration of the right-hand side in Eq. (\ref{32}) is zero
because $\partial_{T}\Delta^{(2)}(T,\mathbf{x}-\mathbf{y})$ vanishes when
$T^{2}-\left\vert \mathbf{x}-\mathbf{y}\right\vert ^{2}\neq0$. Thus, we obtain
the relation%
\begin{equation}
\left[  \int\mathbf{f}_{o}\mathbf{(\mathbf{x})}\cdot\mathbf{\hat{E}\mathbf{(}%
}T\mathbf{\mathbf{,\mathbf{x})}}d^{3}x,~\int\mathbf{a}_{m}(\mathbf{y}%
)\cdot\mathbf{\hat{E}}(0,\mathbf{y})d^{3}y\right]  =0. \label{34}%
\end{equation}
In a similar manner, we can show the following relation using Eq. (\ref{e8})
and Eq. (\ref{e9}).%

\begin{equation}
\left[  \int\mathbf{w}(\mathbf{x})\hat{\varepsilon}\left(  T,\mathbf{x}%
\right)  d^{3}x,~\int\mathbf{a}_{m}(\mathbf{y})\cdot\mathbf{\hat{E}%
}(0,\mathbf{y})d^{3}y\right]  =0. \label{36}%
\end{equation}
From these relations, we can show the commutation relations given by%

\[
\left[  \int\mathbf{w}(\mathbf{x})\hat{\varepsilon}\left(  T,\mathbf{x}%
\right)  d^{3}x,~\hat{M}_{s}^{\dag}\right]  =0,
\]

\[
\left[  \int\mathbf{f}_{o}\mathbf{(\mathbf{x})}\cdot\mathbf{\hat{E}\mathbf{(}%
}T\mathbf{\mathbf{,\mathbf{x})}}d^{3}x,~\hat{M}_{s}^{\dag}\right]  =0
\]
by recalling that $\hat{M}_{s}^{\dag}$ is a function of $\int\mathbf{a}%
_{m}(\mathbf{y})\cdot\mathbf{\hat{E}}(0,\mathbf{y})d^{3}y$, as seen in Eq.
(\ref{81}), Eq. (\ref{e62}), and Eq. (\ref{e63}). Therefore, we can move the
positions of $\hat{M}_{s}^{\dag}$ in the first and second terms of the
right-hand side of Eq. (\ref{31}) to the right, and rewrite $E_{o}$ as%

\begin{align}
&  E_{o}=\langle0|\int\mathbf{w}(\mathbf{x})\hat{\varepsilon}\left(
T,\mathbf{x}\right)  d^{3}x\left(  \sum_{s=\pm}\hat{M}_{s}^{\dag}\hat{M}%
_{s}\right)  |0\rangle\nonumber\\
&  +\theta\langle0|\left(  \int\mathbf{f}_{o}\mathbf{(\mathbf{x})}%
\cdot\mathbf{\hat{E}\mathbf{(}}T\mathbf{\mathbf{,\mathbf{x})}}d^{3}x\right)
\left(  \sum_{s=\pm}s\hat{M}_{s}^{\dag}\hat{M}_{s}\right)  |0\rangle
\nonumber\\
&  +\frac{1}{2}\theta^{2}\int\mathbf{f}_{o}\mathbf{(\mathbf{x})}^{2}%
d^{3}x\langle0|\left(  \sum_{s=\pm}\hat{M}_{s}^{\dag}\hat{M}_{s}\right)
|0\rangle. \label{37}%
\end{align}
By using two relations of the measurement operators given by
\begin{align*}
\sum_{s=\pm}\hat{M}_{s}^{\dag}\hat{M}_{s}  &  =\cos^{2}\hat{G}+\sin^{2}\hat
{G}=1,~\\
\sum_{s=\pm}s\hat{M}_{s}^{\dag}\hat{M}_{s}  &  =\cos^{2}\hat{G}-\sin^{2}%
\hat{G}=\cos\left(  2\hat{G}\right)  ,
\end{align*}
the following expression of $E_{o}$ is obtained.%

\begin{equation}
E_{o}=\langle0|\int\mathbf{w}(\mathbf{x})\hat{\varepsilon}\left(
T,\mathbf{x}\right)  d^{3}x|0\rangle+\theta\eta+\frac{1}{2}\theta^{2}\xi,
\label{38}%
\end{equation}
where $\eta$ and $\xi$ are real constants defined by
\begin{equation}
\eta=\langle0|\left(  \int\mathbf{f}_{o}\mathbf{(\mathbf{x})}\cdot
\mathbf{\hat{E}\mathbf{(}}T\mathbf{\mathbf{,\mathbf{x})}}d^{3}x\right)
\cos\left(  2\hat{G}\right)  |0\rangle, \label{41}%
\end{equation}%
\begin{equation}
\xi=\int\mathbf{f}_{o}\mathbf{(\mathbf{x})}^{2}d^{3}x. \label{90}%
\end{equation}
The first term of the right-hand side of Eq. (\ref{38}) vanishes because we
can transform it into%

\begin{align*}
&  \langle0|\int\mathbf{w}(\mathbf{x})\hat{\varepsilon}\left(  T,\mathbf{x}%
\right)  d^{3}x|0\rangle\\
&  =\langle0|\hat{U}^{\dag}(T)\int\mathbf{w}(\mathbf{x})\hat{\varepsilon
}\left(  0,\mathbf{x}\right)  d^{3}xU(T)|0\rangle\\
&  =\int\mathbf{w}(\mathbf{x})\langle0|\hat{\varepsilon}\left(  \mathbf{x}%
\right)  |0\rangle d^{3}x=0.
\end{align*}
Here, we have used $\hat{U}(T)|0\rangle=|0\rangle$ and Eq. (\ref{72}). The
expression of $\eta$ in Eq. (\ref{41}) can be further simplified by noting that%

\begin{equation}
\cos\left(  2\hat{G}\right)  |0\rangle=\frac{i}{2}\left[  \mathbf{|}\left(
\mathbf{0,}2\mathbf{a}_{m}\right)  \rangle-\mathbf{|}\left(  \mathbf{0,}%
-2\mathbf{a}_{m}\right)  \rangle\right]  . \label{43}%
\end{equation}
By using Eq. (\ref{e22}) and Eq. (\ref{43}), the following relation is
derived.%
\begin{align}
&  \langle0|\mathbf{\hat{E}\mathbf{(}}T\mathbf{\mathbf{,\mathbf{x})}}%
\cos\left(  2\hat{G}\right)  |0\rangle\nonumber\\
&  =-\langle0|\left(  \mathbf{0,}2\mathbf{a}_{m}\right)  \rangle\int
\partial_{T}^{2}\Delta(T,\mathbf{x}-\mathbf{y)a}_{m}(\mathbf{y})d^{3}y,
\label{45}%
\end{align}
where $\Delta\left(  t,\mathbf{x}\right)  $ is a Lorentz invariant function
defined by%
\[
\Delta\left(  t,\mathbf{x}\right)  =\int\frac{d^{3}k}{\left(  2\pi\right)
^{3}|\mathbf{k}|}\cos\left(  \mathbf{k}\cdot\mathbf{x}-|\mathbf{k}|t\right)
.
\]
Contrary to $\Delta^{(1)}\left(  t,\mathbf{x}\right)  $ and $\Delta
^{(2)}\left(  t,\mathbf{x}\right)  $, the function $\Delta\left(
t,\mathbf{x}\right)  $ does not vanish in the off-light-cone region:%

\[
\Delta\left(  t,\mathbf{x}\right)  |_{t^{2}-\mathbf{x}^{2}\neq0}=-\frac
{1}{2\pi^{2}}\frac{1}{t^{2}-\mathbf{x}^{2}}.
\]
The factor $\langle0|\left(  \mathbf{0,}2\mathbf{a}_{m}\right)  \rangle$ is
real and calculated from Eq. (\ref{100}) as%

\[
\langle0\mathbf{|}\left(  \mathbf{0,2a}_{m}\right)  \rangle=\exp\left[
-\int\frac{d^{3}k|\mathbf{k}|}{\left(  2\pi\right)  ^{3}}\left\vert
\mathbf{\tilde{a}}_{m}\left(  \mathbf{k}\right)  \right\vert ^{2}\right]  ,
\]
where $\mathbf{\tilde{a}}_{m}\left(  \mathbf{k}\right)  =\int\mathbf{a}%
_{m}(\mathbf{x})e^{-i\mathbf{k}\cdot\mathbf{x}}d^{3}x$. Substituting Eq.
(\ref{45}) into Eq. (\ref{41})\ gives the final expression of $\eta$ as%
\begin{equation}
\eta=\langle0|\left(  \mathbf{0,}2\mathbf{a}_{m}\right)  \rangle\int
\int\mathbf{\partial}_{T}^{2}\Delta(T,\mathbf{x}-\mathbf{y)f}_{o}%
\mathbf{(\mathbf{x})}\cdot\mathbf{a}_{m}(\mathbf{y})d^{3}xd^{3}y. \label{e100}%
\end{equation}
Minimization of $E_{o}$ with respect to $\theta$ is achieved by taking the
parameter $\theta$ as%

\begin{equation}
\theta=-\frac{\eta}{\xi}. \label{64}%
\end{equation}
Substituting the value of $\theta$ into Eq. (\ref{38}) yields the following
expression of $E_{o}$.%

\begin{equation}
E_{o}=-\frac{\eta^{2}}{2\xi}. \label{40}%
\end{equation}
Substituting Eq. (\ref{90}) and Eq. (\ref{e100}) into Eq. (\ref{40}), we
obtain our main result in this section:%

\begin{equation}
E_{o}=-D_{q}\frac{\left[  \int\int\mathbf{\partial}_{T}^{2}\Delta
(T,\mathbf{x}-\mathbf{y)f}_{o}\mathbf{(\mathbf{x})}\cdot\mathbf{a}%
_{m}(\mathbf{y})d^{3}xd^{3}y\right]  ^{2}}{2\int\mathbf{f}_{o}%
\mathbf{(\mathbf{x})}^{2}d^{3}x}, \label{e200}%
\end{equation}
where $D_{q}\,$\ is an exponential damping factor with respect to
$\mathbf{a}_{m}$ such that
\begin{equation}
D_{q}=\left\vert \langle0|\left(  \mathbf{0,}2\mathbf{a}_{m}\right)
\rangle\right\vert ^{2}=\exp\left[  -2\int\frac{d^{3}k|\mathbf{k}|}{\left(
2\pi\right)  ^{3}}\left\vert \mathbf{\tilde{a}}_{m}\left(  \mathbf{k}\right)
\right\vert ^{2}\right]  . \label{e202}%
\end{equation}
Because $\xi>0$, as seen in Eq. (\ref{90}), it is of importance that $E_{o}$
is negative in Eq. (\ref{40}):%
\[
E_{o}=-\left\vert E_{o}\right\vert <0.
\]
Respecting local energy conservation, this result implies that positive energy
$+|E_{o}|$ moves from the field to external systems during the displacement
operation $\hat{U}_{s}$ because the energy of the field around $V_{m}$ is zero
before the operation. Nonnegativity of the total energy of the field ensures
that $|E_{o}|$ is smaller than $E_{m}$, as discussed in reference
\cite{hotta1}. The wave packets with negative energy begin to chase after the
positive-energy wave packets generated by the first measurement. Figure 2
shows the energy extraction process (step (2)). The outside doughnut-shaped
region represents the propagating wave packet generated by the measurement in
step (1). The displacement operation $\hat{U}_{s}$ is performed in the
measurement area $V_{m}$ denoted by the circle in the figure. In Figure 3, the
average energy density is plotted immediately before the displacement
operation at $t=T-0$. We have zero energy around the area where $\hat{U}_{s}$
is performed. In Figure 4, the average energy density after the operation is
plotted. Negative energy density appears around $V_{m}$. In conclusion, a part
of the escaped energy is effectively retrieved to the measurement area by this
QET protocol.

In this discrete-variable case with the spin probe, the amount of extracted
energy $\left\vert E_{o}\right\vert $ in Eq. (\ref{e200}) is suppressed by the
exponential damping factor $D_{q}$ in Eq. (\ref{e202}) when the magnitude of
$\mathbf{a}_{m}$ and the infused energy $E_{m}$ increase. This suppression
becomes power damping for a continuous-variable teleportation with a probe
harmonic oscillator, as seen in the next section.

\bigskip

\section{Continuous-Variable Teleportation}

\ \newline

In this section, we analyze a QET protocol with a continuous variable.\ The
essential part of the protocol is the same as that discussed in section 3
except the probe system and the operation dependent on the measurement result.
As the probe to measure the quantum fluctuation of electric field, let us
consider a harmonic oscillator. The free Hamiltonian is given by%

\begin{equation}
\hat{H}_{ho}=\frac{1}{2}\hat{p}^{2}+2\hat{q}^{2}, \label{200}%
\end{equation}
where $\hat{q}$ is the position operator and $\hat{p}$, the momentum operator.
This oscillator couples with the electric field by the measurement interaction
such that
\[
\hat{H}_{m}^{\prime}(t)=g(t)\hat{p}\hat{G},
\]
where $g(t)$ is the time-dependent factor in Eq. (\ref{e1}) and $\hat{G}$, the
Hermitian operator in Eq. (\ref{81}). The initial state of the oscillator is
the ground state $|g\rangle$ of $H_{ho}$. Introducing an eigenstate
$|q\rangle~$\ of $\hat{q}$ with eigenvalue $q$, the ground state is described
by
\[
\langle q|g\rangle=\left(  \frac{2}{\pi}\right)  ^{1/4}\exp\left[
-q^{2}\right]  .
\]
After the measurement interaction is switched off, we measure the position
$\hat{q}$. The measurement operator for measurement result $q$ is computed as%

\[
\hat{M}_{q}=\langle q|\exp\left[  -i\hat{p}\hat{G}\right]  |g\rangle.
\]
More explicitly, $\hat{M}_{q}$ is written as
\begin{equation}
\hat{M}_{q}=\left(  \frac{2}{\pi}\right)  ^{1/4}\exp\left[  -\left(  q-\hat
{G}\right)  ^{2}\right]  . \label{213}%
\end{equation}
The operator $\hat{M}_{q}$ satisfies the following relations.
\begin{align}
\int_{-\infty}^{\infty}\hat{M}_{q}^{\dag}\hat{M}_{q}dq  &  =1,\label{201}\\
\int_{-\infty}^{\infty}q\hat{M}_{q}^{\dag}\hat{M}_{q}dq  &  =\hat
{G},\label{202}\\
\int_{-\infty}^{\infty}q^{2}\hat{M}_{q}^{\dag}\hat{M}_{q}dq  &  =\hat{G}%
^{2}+\frac{1}{4}. \label{203}%
\end{align}
The state after the measurement is given by%
\[
\hat{\rho}_{m}^{\prime}=\int\hat{M}_{q}|0\rangle\langle0|\hat{M}_{q}^{\dag
}dq.
\]
In this measurement, energy is infused into the field as given by the protocol
described in section 3. The expectation value of the infused energy density is
evaluated as%
\begin{equation}
\left\langle \hat{\varepsilon}(t,\mathbf{x})\right\rangle =\operatorname*{Tr}%
\left[  \hat{\rho}_{m}^{\prime}\hat{\varepsilon}(t,\mathbf{x})\right]
=\int\langle0|\hat{M}_{q}^{\dag}\hat{\varepsilon}(t,\mathbf{x})\hat{M}%
_{q}|0\rangle dq, \label{205}%
\end{equation}
where $\hat{\varepsilon}(t,\mathbf{x})=\hat{U}^{\dag}(t)\hat{\varepsilon
}\left(  \mathbf{x}\right)  \hat{U}(t)$. The value $\left\langle
\hat{\varepsilon}(t,\mathbf{x})\right\rangle $ can be calculated by using the
following integral formula.%

\begin{equation}
\exp\left(  -q^{2}\right)  =\frac{1}{\sqrt{4\pi}}\int e^{-\frac{p^{2}}{4}%
-ipq}dp. \label{206}%
\end{equation}
From Eq. (\ref{206}), the following relation is derived.%
\begin{align*}
\hat{M}_{q}|0\rangle &  =\left(  \frac{2}{\pi}\right)  ^{1/4}\exp\left[
-\left(  q-\hat{G}\right)  ^{2}\right]  |0\rangle\\
&  =\left(  \frac{2}{\pi}\right)  ^{1/4}\frac{1}{\sqrt{4\pi}}\int
dpe^{-\frac{p^{2}}{4}-ipq}\exp\left[  ip\hat{G}\right]  |0\rangle.
\end{align*}
From Eq. (\ref{e4}), $\hat{M}_{q}|0\rangle$ is rewritten as%

\begin{equation}
\hat{M}_{q}|0\rangle=\left(  \frac{2}{\pi}\right)  ^{1/4}\frac{1}{\sqrt{4\pi}%
}\int dpe^{-\frac{p^{2}}{4}-ip\left(  q-\frac{\pi}{4}\right)  }|\left(
\mathbf{0},p\mathbf{a}_{m}\right)  \rangle. \label{207}%
\end{equation}
Substituting Eq. (\ref{207}) into Eq. (\ref{205}) yields the following manipulation:%

\begin{align}
&  \left\langle \hat{\varepsilon}(t,\mathbf{x})\right\rangle \nonumber\\
&  =\frac{1}{\sqrt{2\pi}}\int\int e^{-\frac{\bar{p}^{2}}{4}-\frac{p^{2}}{4}%
}\langle\left(  \mathbf{0},\bar{p}\mathbf{a}_{m}\right)  |\hat{\varepsilon
}(t,\mathbf{x})|\left(  \mathbf{0},p\mathbf{a}_{m}\right)  \rangle\left(  \int
e^{i\left(  \bar{p}-p\right)  \left(  q-\frac{\pi}{4}\right)  }\frac{dq}{2\pi
}\right)  d\bar{p}dp\nonumber\\
&  =\frac{1}{\sqrt{2\pi}}\int dpe^{-\frac{p^{2}}{2}}\langle\left(
\mathbf{0},p\mathbf{a}_{m}\right)  |\hat{\varepsilon}(t,\mathbf{x})|\left(
\mathbf{0},p\mathbf{a}_{m}\right)  \rangle\nonumber\\
&  =\frac{1}{\sqrt{2\pi}}\int p^{2}e^{-\frac{p^{2}}{2}}dp\langle\left(
\mathbf{0},\mathbf{a}_{m}\right)  |\hat{\varepsilon}(t,\mathbf{x})|\left(
\mathbf{0},\mathbf{a}_{m}\right)  \rangle\nonumber\\
&  =\langle\left(  \mathbf{0},\mathbf{a}_{m}\right)  |\hat{\varepsilon
}(t,\mathbf{x})|\left(  \mathbf{0},\mathbf{a}_{m}\right)  \rangle, \label{208}%
\end{align}
where we have used $\langle\left(  \mathbf{0},p\mathbf{a}_{m}\right)
|\hat{\varepsilon}(t,\mathbf{x})|\left(  \mathbf{0},p\mathbf{a}_{m}\right)
\rangle=p^{2}\langle\left(  \mathbf{0},\mathbf{a}_{m}\right)  |\hat
{\varepsilon}(t,\mathbf{x})|\left(  \mathbf{0},\mathbf{a}_{m}\right)  \rangle
$. From Eq. (\ref{208}), we can show that the input energy $\int_{-\infty
}^{\infty}\left\langle \hat{\varepsilon}(0,\mathbf{x})\right\rangle d^{3}x$ in
the measurement process is the same as that of the discrete-variable case
described by Eq. (\ref{e7}) of section 3. Besides, in the same manner as the
derivation of Eq. (\ref{e16}), it is possible from Eq. (\ref{208}) to show the
equation
\[
\langle\hat{\varepsilon}(t,\mathbf{x})\rangle=\frac{1}{2}\left[  \mathbf{\Pi
}(t,\mathbf{x)}^{2}+\mathbf{b}(t,\mathbf{x)}^{2}\right]  ,
\]
where $\mathbf{b}(t,\mathbf{x)}$ and $\mathbf{\Pi}(t,\mathbf{x)}$ are given by
Eq. (\ref{e10}) and Eq. (\ref{e12}), respectively.

After the wave packets generated by the measurement escape from the
measurement area, we execute at\thinspace$t=T$, when the field has zero energy
inside $V_{m}$, a local unitary operation dependent of the measurement result
$q$ given by%
\begin{equation}
\hat{U}_{q}=\exp\left[  iq\theta^{\prime}\int\mathbf{f}_{o}(\mathbf{x}%
)\cdot\mathbf{\hat{A}}(\mathbf{x})d^{3}x\right]  , \label{215}%
\end{equation}
which is analogous to Eq. (\ref{57}). The parameter $\theta^{\prime}$ is real
and fixed later. Then, the energy around the measurement area is evaluated as
\[
E_{o}^{\prime}=\int dq\langle0|\hat{M}_{q}^{\dag}\hat{U}^{\dag}(T)\hat{U}%
_{q}^{\dag}\hat{H}_{o}\hat{U}_{q}\hat{U}(T)\hat{M}_{q}|0\rangle.
\]
By using Eq. (\ref{201})-Eq. (\ref{203}) and%

\begin{align*}
&  \hat{U}^{\dag}(T)\hat{U}_{q}^{\dag}\hat{H}_{o}\hat{U}_{q}\hat{U}(T)\\
&  =\int\mathbf{w}(\mathbf{x})\hat{\varepsilon}\left(  T,\mathbf{x}\right)
d^{3}x\\
&  +\theta^{\prime}q\int\mathbf{f}_{o}\mathbf{(\mathbf{x})}\cdot
\mathbf{\hat{E}\mathbf{(}}T\mathbf{\mathbf{,\mathbf{x})}}d^{3}x+\frac{1}%
{2}\theta^{\prime2}q^{2}\int\mathbf{f}_{o}\mathbf{(\mathbf{x})}^{2}d^{3}x,
\end{align*}
the average energy $E_{o}^{\prime}$ is rewritten, in a manner similar to that
in section 3, as
\[
E_{o}^{\prime}=\theta^{\prime}\eta^{\prime}+\frac{\xi}{2}\theta^{\prime
2}\langle0|\left(  \hat{G}^{2}+\frac{1}{4}\right)  |0\rangle,
\]
where $\xi$ is given by Eq. (\ref{90}) and $\eta^{\prime}$ is a real constant
that is evaluated as
\begin{align}
\eta^{\prime}  &  =\langle0|\left(  \int\mathbf{f}_{o}\mathbf{(\mathbf{x}%
)\cdot\hat{E}\mathbf{(}}T\mathbf{\mathbf{,\mathbf{x})}}d^{3}x\right)  \hat
{G}|0\rangle\nonumber\\
&  =-\langle0|\left(  \int\mathbf{f}_{o}\mathbf{(\mathbf{x})\cdot\hat
{E}\mathbf{(}}T\mathbf{\mathbf{,\mathbf{x})}}d^{3}x\right)  \left(
\int\mathbf{a}_{m}(\mathbf{y})\cdot\mathbf{\hat{E}}(\mathbf{y})d^{3}y\right)
|0\rangle\nonumber\\
&  =\frac{1}{2}\int\int\mathbf{\partial}_{T}^{2}\Delta(T,\mathbf{x}%
-\mathbf{y)f}_{o}\mathbf{(x)}\cdot\mathbf{a}_{m}(\mathbf{y})d^{3}xd^{3}y.
\label{500}%
\end{align}
By fixing the parameter $\theta^{\prime}$ as
\begin{equation}
\theta^{\prime}=-\frac{\eta^{\prime}}{\xi\langle0|\left(  \hat{G}^{2}+\frac
{1}{4}\right)  |0\rangle}, \label{217}%
\end{equation}
the minimization of $E_{o}^{\prime}$ with respect to $\theta^{\prime}$ is
attained as%
\[
E_{o}^{\prime}=-\frac{\eta^{\prime2}}{2\xi\langle0|\left(  \hat{G}^{2}%
+\frac{1}{4}\right)  |0\rangle}.
\]
By substituting Eq. (\ref{90})\ and Eq. (\ref{500}) into the above equation,
we obtain the final expression of $E_{o}^{\prime}$ as
\begin{equation}
E_{o}^{\prime}=-D_{ho}\frac{\left[  \int\int\mathbf{\partial}_{T}^{2}%
\Delta(T,\mathbf{x}-\mathbf{y)f}_{o}\mathbf{(x)}\cdot\mathbf{a}_{m}%
(\mathbf{y})d^{3}xd^{3}y\right]  ^{2}}{2\int\mathbf{f}_{o}\mathbf{(\mathbf{x}%
)}^{2}d^{3}x}, \label{211}%
\end{equation}
where $D_{ho}$ is given by%
\begin{equation}
D_{ho}=\left[  \langle0|\left(  4\hat{G}^{2}+1\right)  |0\rangle\right]
^{-1}=\left[  1+\frac{\pi^{2}}{4}+2\int|\mathbf{k}|\left\vert \mathbf{\tilde
{a}}_{m}\left(  \mathbf{k}\right)  \right\vert ^{2}\frac{d^{3}k}{\left(
2\pi\right)  ^{3}}\right]  ^{-1}. \label{212}%
\end{equation}
It is noticed that the form of $E_{o}^{\prime}$ in Eq. (\ref{211}) is the same
as that of $E_{o}$ in Eq. (\ref{e200}), except for the form of damping factor
$D_{q}$. Unlike the exponential damping of $D_{q}$, the suppression factor
$D_{ho}$ is a power damping factor with respect to $\left\vert \mathbf{\tilde
{a}}_{m}\left(  \mathbf{k}\right)  \right\vert $. Because of this weak
damping, the extracted energy $\left\vert E_{o}^{\prime}\right\vert $ by this
protocol does not vanish even for a large amplitude limit with $\left\vert
\mathbf{\tilde{a}}_{m}\left(  \mathbf{k}\right)  \right\vert \rightarrow
\infty$ as%

\[
\left\vert E_{o}^{\prime}\right\vert \sim\frac{\left\vert \int|\mathbf{k}%
_{1}|\cos\left(  |\mathbf{k}_{1}|T\right)  \mathbf{n}_{m}\left(
\mathbf{k}_{1}\right)  ^{\ast}\cdot\mathbf{\tilde{f}}_{o}\left(
\mathbf{k}_{1}\right)  \frac{d^{3}k_{1}}{\left(  2\pi\right)  ^{3}}\right\vert
^{2}}{4\left(  \int\left\vert \mathbf{\tilde{f}}_{o}\left(  \mathbf{k}%
_{2}\right)  \right\vert ^{2}\frac{d^{3}k_{2}}{\left(  2\pi\right)  ^{3}%
}\right)  },
\]
where $\mathbf{\tilde{f}}_{o}\left(  \mathbf{k}\right)  =\int\mathbf{f}%
_{o}(\mathbf{x})e^{-i\mathbf{k}\cdot\mathbf{x}}d^{3}x$ and $\mathbf{n}%
_{m}\left(  \mathbf{k}\right)  $ is a rescaled amplitude given by%
\[
\mathbf{n}_{m}\left(  \mathbf{k}\right)  =\frac{\mathbf{\tilde{a}}_{m}\left(
\mathbf{k}\right)  }{\sqrt{\int|\mathbf{k}^{\prime}|\left\vert \mathbf{\tilde
{a}}_{m}\left(  \mathbf{k}^{\prime}\right)  \right\vert ^{2}\frac
{d^{3}k^{\prime}}{\left(  2\pi\right)  ^{3}}}}.
\]
As a conclusion, we can say that obtaining more information of the fluctuation
leads to teleporting more energy. It is an important question what
measurements and operations attain the maximum transporation rate of energy in
the QET mechanism, however, remains unsolved yet.

\section{Summary and Discussion}

\ \newline

We have analyzed in detail two protocols of QET for the electromagnetic field
and shown that a part of the lost energy in the measurement can be retrieved
by use of the measurement result. The amount of energy infused by the
measurement is the same in both the cases and it is given by Eq. (\ref{e7}).
For the discrete-variable case, the measurement with the operator $\hat{M}%
_{s}$ given by Eq. (\ref{e62}) and Eq. (\ref{e63}) is performed. The amount of
retrieval energy $\left\vert E_{o}\right\vert $ is given by Eq. (\ref{e200})
for the displacement operator in Eq. (\ref{57}) with $\theta$ fixed in Eq.
(\ref{64}). For the continuous-variable case, the measurement with the
operator $\hat{M}_{s}$ given by Eq. (\ref{213}) is performed. The amount of
energy retrieved $\left\vert E_{o}^{\prime}\right\vert $ is given by Eq.
(\ref{211}) for the displacement operator in Eq. (\ref{215}) with $\theta$
fixed in Eq. (\ref{217}). For large amplitude $\left\vert \mathbf{\tilde{a}%
}_{m}\left(  \mathbf{k}\right)  \right\vert $ and large energy input $E_{m}$,
the continuous-variable teleportation is found to be more preferable than the
discrete-variable teleportation. In the discrete-variable case, the amount of
extracted energy is suppressed by the exponential damping factor $D_{q}$ in
Eq. (\ref{e202}) when the energy infused by the measurement increases. For the
continuous-variable case, the suppression factor becomes the power damping
factor $D_{ho}$ in Eq. (\ref{212}). Therefore, it is concluded that obtaining
more information about the quantum fluctuation leads to teleporting more energy.

In future QET experiments, the separation between $V_{m}$ and the region of
the escaping wave packet with positive energy should not be so large to
extract an observable amount of energy. To see this, let us take a
large-separation limit as $T\gg\left\vert \mathbf{x-y}\right\vert $ for
$\mathbf{x,y\in V}_{m}$. Then, the amount of teleported energy decays as%

\[
\left\vert E_{o}\right\vert \propto\frac{1}{T^{12}},
\]
in both the protocols. This rapid decay becomes one of the serious obstacles
to observing the extraction of energy by QET. Therefore, for the best
implementation of the QET in 1+3 dimensions, $T\,$\ should be of the same
order as the measurement area size. From this viewpoint, QET protocols are
more attractive for physical systems described effectively by 1+1 dimensional
massless field models that have slower-decay properties ($E_{o}$ $\propto
T^{-4}$) of the transported energy for a large separation $T$, as discussed in
references \cite{hotta1}-\cite{hotta2}.

\bigskip

\textbf{Acknowledgments}\newline

This research was partially supported by the SCOPE project of the MIA and the
Ministry of Education, Science, Sports and Culture of Japan, No. 21244007.

\bigskip

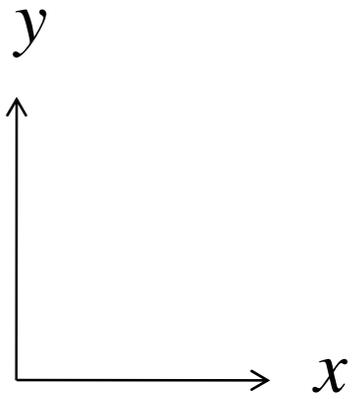
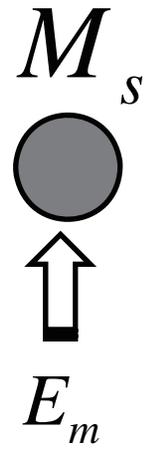
Figure 1: The first schematic diagram of QET in the $xy$-plane slice. The
local measurement is performed with the infusion of positive energy $E_{m}$ to
the field and the measurement result $s$ is obtained. The measurement area
$V_{m}$ is represented by a circle in the plane. A positive-energy wave packet
is generated in the field system and it escapes to spatial infinity at the
velocity of light.

\bigskip

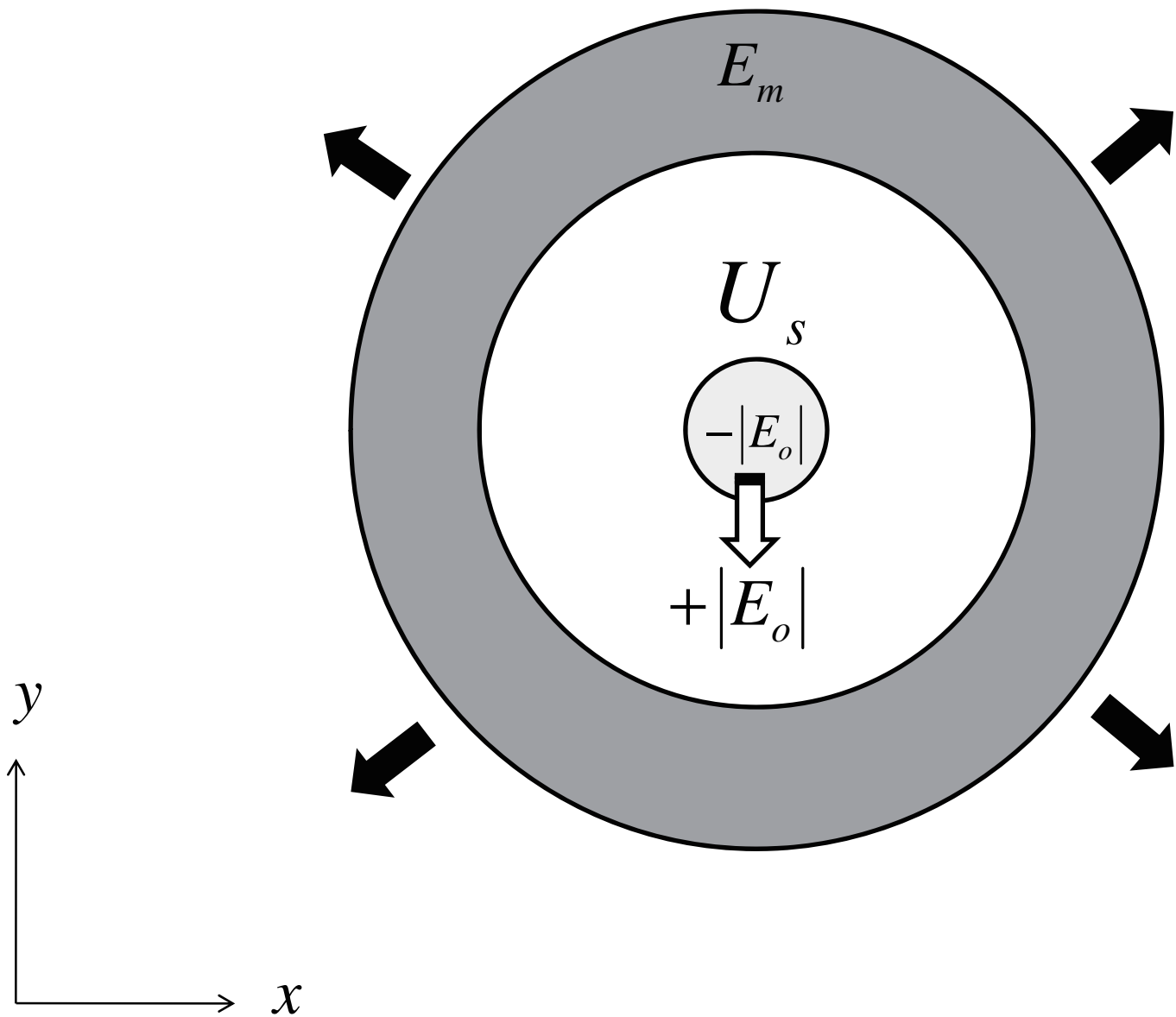
Figure 2: The second schematic diagram of QET in the $xy$-plane slice. The
outside doughnut-shaped region represents the propagating wave packet
generated by the measurement. By using the measurement result $s$, the
displacement operation $\hat{U}_{s}$ is carried out in the measurement area
$V_{m}$ denoted by the circle. After the operation, a wave packet with
negative energy $-\left\vert E_{o}\right\vert $ is generated, accompanying the
extraction of positive energy $+\left\vert E_{o}\right\vert $ to external systems.

\bigskip

\bigskip

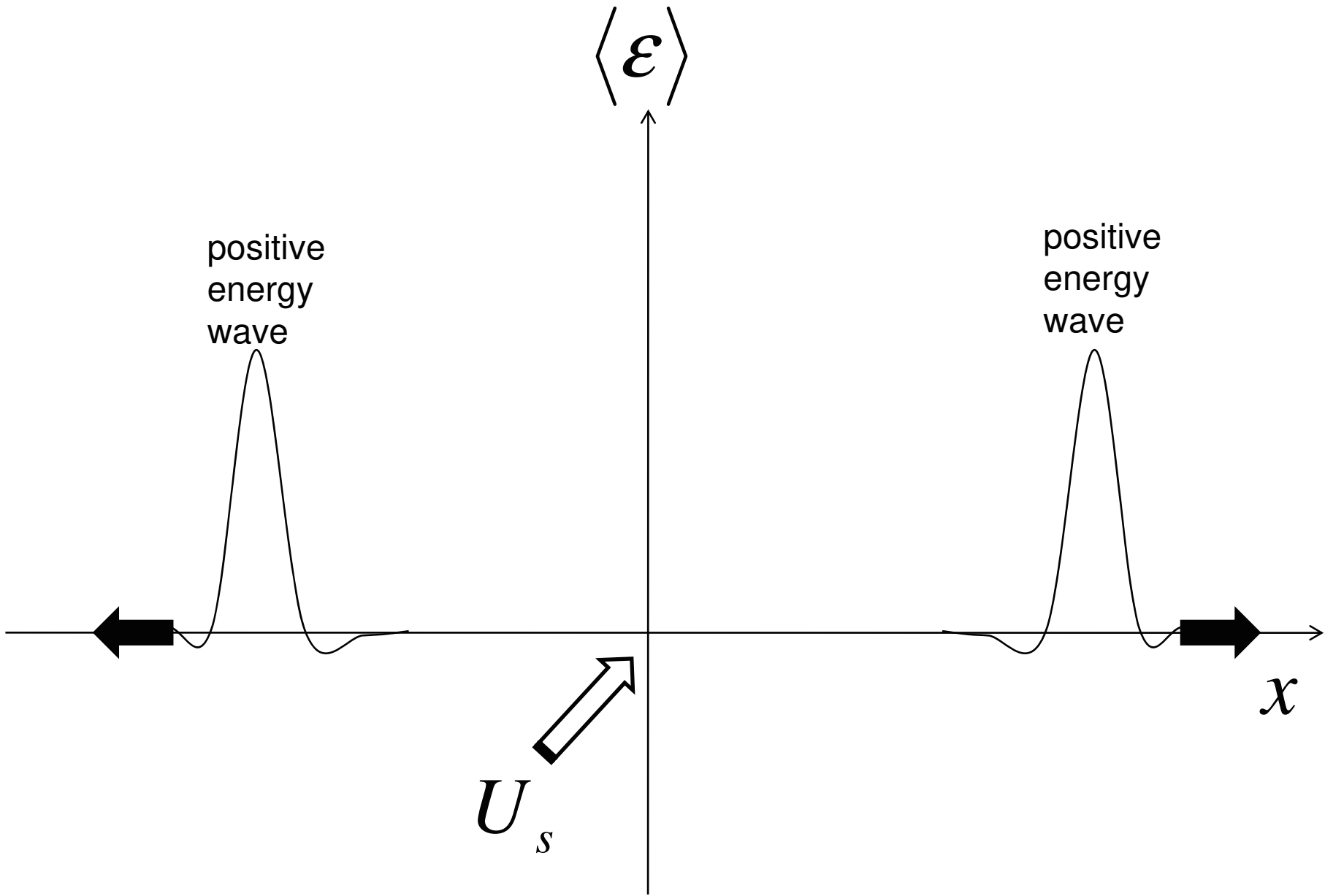
Figure 3: The average energy density is schematically plotted immediately
before the displacement operation at $t=T-0$. The positive-energy wave packet
generated by the measurement has already escaped from the measurement area
where $\hat{U}_{s}$ is performed.

\bigskip

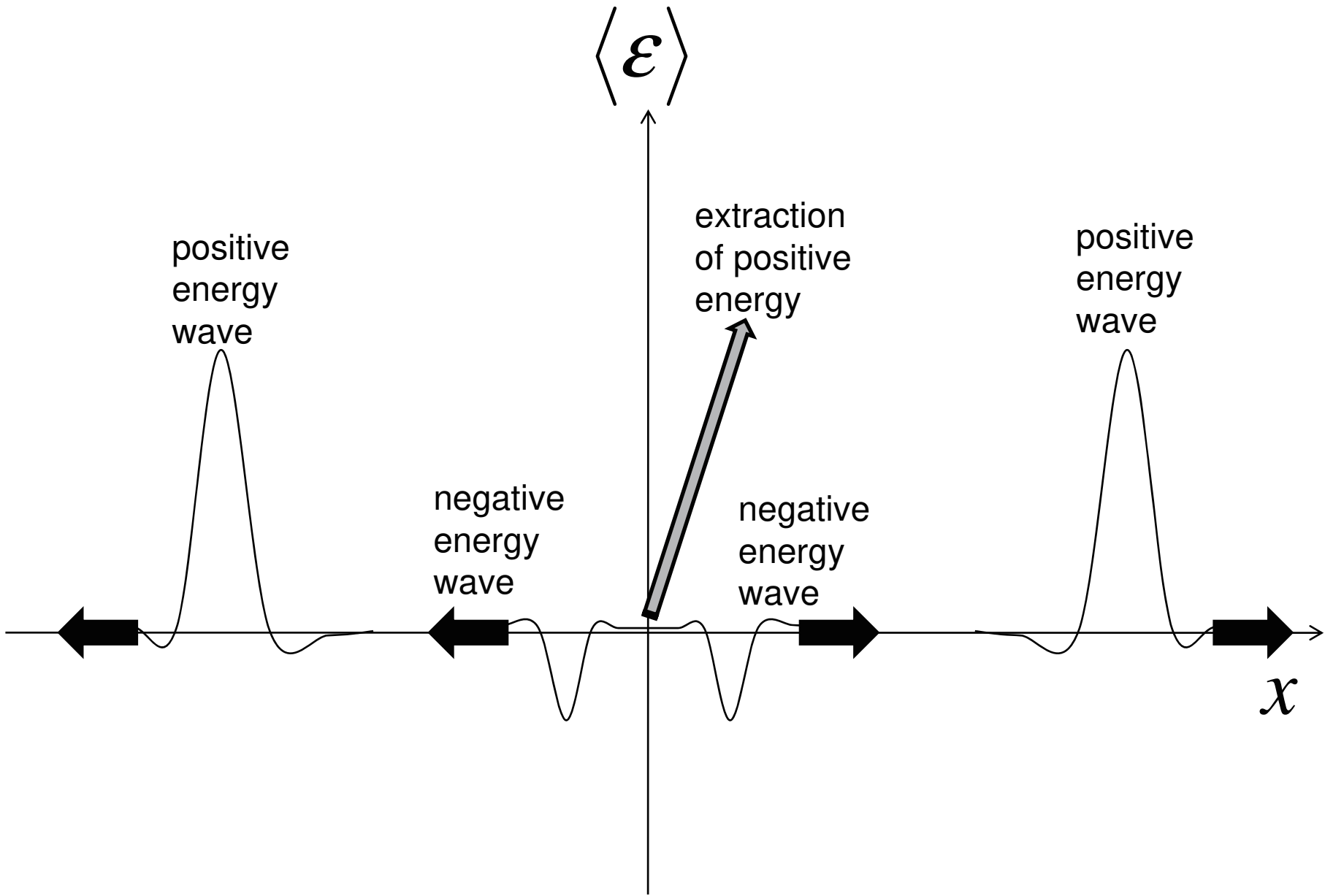
Figure 4: The average energy density is schematically plotted after the
operation. The wave packet with negative energy $-\left\vert E_{o}\right\vert
$ is generated by the extraction of positive energy $+\left\vert
E_{o}\right\vert $ to external systems.

\end{document}